\documentclass[12pt, a4]{elsarticle}

\usepackage{graphicx}
\usepackage{dcolumn}
\usepackage{bm}

\usepackage[utf8]{inputenc}
\usepackage[T1]{fontenc}
\usepackage{mathptmx}
\usepackage{etoolbox}
\usepackage{lineno,setspace}
\usepackage{stackrel,amssymb,amsmath}
\usepackage{pdflscape}

\newtheorem{theorem}{\small{Theorem}} 
\newtheorem{corollary}{\small{\em Corollary}} 

\journal{Measurement}
\begin{document}

\begin{frontmatter}

\title{A new symmetry-based extraction method of Schottky diode parameters from resistance-compensated I-V characteristics}

\author[label1]{Ocaya R.O.\corref{cor1}}
\address[label1]{Department of Physics, Univ. of the Free State, P. Bag X13, Phuthaditjhaba 9866, South Africa}
\cortext[cor1]{Corresponding author}
\ead{ocayaro@ufs.ac.za}

\author[label2]{Yakuphano\u glu  F}
\address[label2]{Department of Physics, Faculty of Science, Firat University, Elazig, Turkey}

\begin{abstract}
We present a novel resistance-compensated I-V method to extract the series resistance, ideality factor, barrier height and built-in potential of a metal-semiconductor diode. We show that a reduced equation arises from a unique but hitherto unreported symmetry in the Schottky equation when it is written as an ordinary differential equation. In spite of the intense mathematical justification, we show how this new equation is directly applicable to an empirical data set through a simple algorithm. We test the method on a new Al/p-Si/Bi$_2$Se$_3$/Al Schottky diode and compare it with the Cheung-Cheung method on the same data. The series resistance was found to change exponentially with applied bias with a rate constant that depends on the incident illumination. The barrier height decreased with bias but was independent of the incident illumination. The trends in the results of the method agree strongly with the literature and may yield more accurate diode parameters compared to other electrical methods.
\end{abstract}


\end{frontmatter}

\section{Introduction}

The metal-semiconductor (MS) or Schottky diode arises when specific metals form two junction interfaces with a semiconductor. One junction is rectifying while the second serves as an Ohmic, ideally zero-resistance electrical lead-out from the device. Since all real solid-state devices are made using metal-semiconductor interfaces they are potentially Schottky junctions, thus accounting for the expansive literature and the current flurry of research interest in new semiconductors and suitable interface metallization. Schottky diodes are as many and varied as there are applications for solid-state devices. They find use in fast-recovery rectifiers, all kinds of RF, thermal and opto-electronic detectors, transistors, and innumerable other applications. For new experimental semiconductor materials especially it is necessary to quantify the intrinsic parameters of both the material and the final MS device. There are many methods for this, such as optical spectrometry, impedance and electrical measurements, each with its advantages and limitations \cite{norde1979,nicollian1982mos,cheung1986extraction,zhang2005effects}. Within the electrical methods, for instance, there is much method variance as each approach has its own complexity, utility, and weaknesses. The difficulty of the electrical parameter extraction is the possible interdependence of the parameters of interest. The problem is compounded by the cyclical expression of the current. The barrier height, series resistance, and ideality factor, specifically, remain important but challenging parameters to extract through electrical measurements. The series resistance features significantly in the electrical determination of the other parameters, such as barrier height and interface state density \cite{Guler2009,lapa2019effect}. The literature provides ample evidence of large variances in the determined series resistance calculated using different methods on the same data. Many workers have suggested methods to compensate for the series resistance \cite{turut1996bias,gromov1994modified,manifacier1988schottky}. Series resistance has a practical bearing since it limits the useful power of the device through the short-circuit current when the device is used in its active region. This is disadvantageous in photovoltaics where the output power of the device must be characterised accurately. Series resistance therefore plays an important but often understated role. 
 
In this article, we apply group theory to the challenging problem of parameter extraction from the Schottky equation in the presence of appreciable series resistance. The idea arose when considering Noether's first theorem \cite{noether1971} which states that when conservative forces act on a system with a differentiable symmetry then the action has a corresponding conservation law. Our starting point was, therefore, the search for possible symmetry in the MS diode I-V characteristics. Having established the existence of a symmetry, we then proved that it leads to the instantaneous series resistance. The series resistance can then be used to compensate the characteristics, and leads to a more reliable barrier height and ideality factor calculation. Symmetry analysis has been around for a long time \cite{ibragimov1992group} but its apparent complexity has kept it in the realm of abstract mathematics. Lie groups are sets of differentiable manifolds together with their admissible algebras. Their structures and properties have been studied intensely over the years but they remain too abstract for many, with few clear examples connecting them to the applied sciences. Yet symmetry analysis has methods that unify many of the methods of solving ordinary differential equations (ODEs). Its main premise is that if a non-trivial solution of a given ODE is known, then the other solutions can be found by continuously varying an infinitesimal parameter. Based on this, we propose a novel method to simplify the analysis of the Schottky equation by rigorously addressing the series resistance. We obtain a surprisingly simple result that leads to a straightforward algorithm to apply the method on empirical data.

\section{The hypothesis}

The diode current current $I$ in a Schottky diode in the presence of appreciable diode series resistance $R_s$ is cyclically dependent on itself, the applied voltage $V$, the absolute diode temperature $T$, effective barrier height $\phi$, the ideality factor $n$  \cite{rhoderick1988metal,durmus2011,salambdaam1996series}. That is,
\begin{eqnarray}
    \nonumber I&=&AA^*T^2\exp\Big(-\frac{q\phi}{kT}\Big)\exp\Big(\frac{qV-IR_s}{nkT}\Big)\\
    &&\times \Big\{1-\exp\Big(-\frac{qV-IR_s}{kT}\Big)\Big\},
    \label{eqn:schottky}
\end{eqnarray}
where $q$ is electronic charge, $k$ is the Boltzmann constant, $A$ is the diode area, $A^*$ is the Richardson constant.
Eq. \ref{eqn:schottky} is difficult to solve explicitly because of $R_s$. Existing methods estimate $\phi$ and $n$ by using simplifying approximations. A commonly used assumption is that $qV$ < $3kT$ such that $n$, $R_s$ are estimated through $\ln (I)$ and $\ln (V)$ functions. This limits the method to the forward low-bias region. The barrier height $\phi$ is known to depend on applied bias and many simplifying assumptions and treatments of this variation are also described in the literature \cite{werner1988,rhoderick1988metal,chand1997,yildiz2008,chowdhury2019,khanna2011}.

For brevity Eq. \ref{eqn:schottky} can be written in a more general form, where barrier height varies with applied voltage, i.e. $\phi$=$\phi(V)$:
\begin{equation}
    y = (1/s)e^{[-b\phi(x)+c(x-yr)]}\Big\{1-e^{-b(x-yr)}\Big\},
    \label{eqn:schottky1a}
\end{equation}
where $x$=$V$, $y$=$I$, $b$=$q/kT$, $c$=$b/n$, $r$=$R_s$, and ($1/s$)=$AA^*T^2$. If $\phi$ is assumed to be constant then Eq. \ref{eqn:schottky1a} takes the simpler form:
\begin{equation}
    y = (1/p)e^{c(x-yr)}\Big\{1-e^{-b(x-yr)}\Big\},
    \label{eqn:schottky1}
\end{equation}
where $s$=$p\exp(-b\Phi_0)$, where $\phi$=$\Phi_0$ is the zero-bias barrier height. This parameter is itself of interest. There are, therefore, two possible broad treatments depending on $\phi$. In the first case $\phi$=$\phi_o$ = constant, whereas it is not in the second case. To illustrate the general arguments without complex equations, we will begin with the first case i.e at Eq. \ref{eqn:schottky1} where $\{p,b,c,r\}$ are constants. Then, we will then return to Eq. \ref{eqn:schottky1a}, which is the second case we shall consider.

\subsection{Case I - Constant barrier height}

\begin{figure}
\centering
\includegraphics{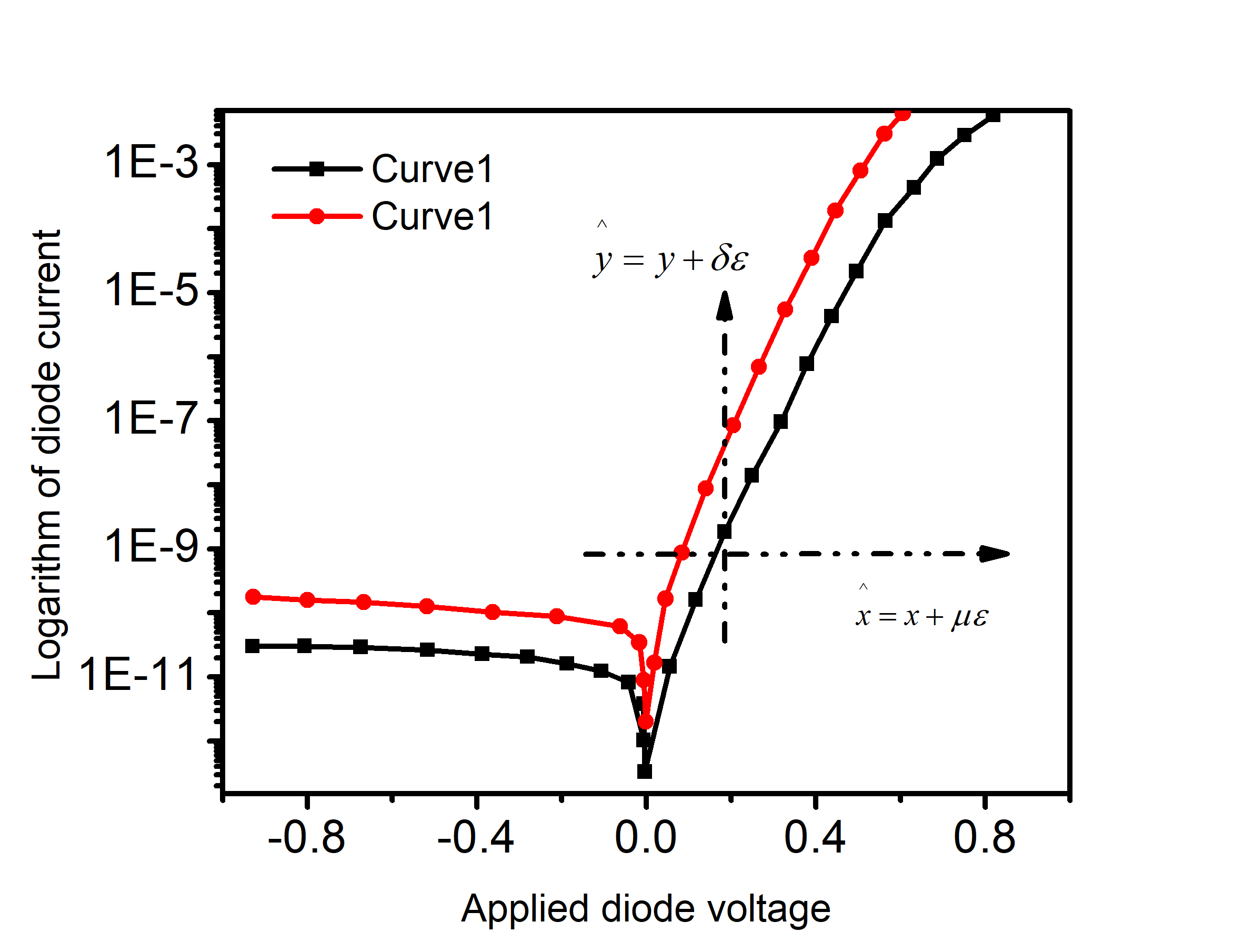}
\caption{\label{fig:symmetry} Typical parametric plot of current-voltage characteristics of a Schottky diode with series resistance. The tanslational symmetry is apparent from Curve 1 to Curve 2 along both $x$ and $y$ axes. There is possibly a scaling symmetry along $y$.}
\end{figure}

Fig. \ref{fig:symmetry} shows two traces, Curve 1 and Curve 2, that are typical of I-V characteristics of MS diodes. These curves can be taken as continuously differentiable solutions of some well-defined ODE. The uniqueness of each solution will then depend on the resistance, ambient temperature, doping density and so on. The visual similarity of the two curves suggests that there may exist a well-defined mapping that takes one solution to the other interchangeably. If so, then the two curves meet the necessary conditions for the existence of a symmetry group. The first condition is closure, since there would be a well-defined set of solution curves. The second condition is the existence of a mapping that takes one solution (a unique ($x$,$y$) point) of the ODE to other unique points. The path of the mapping is an orbit. Finally, the mapping should be reversible. The following theorem conveys the idea.

\begin{theorem}\label{thm:first}
For any Schottky diode described by Eq. \ref{eqn:schottky1} there exists a translational, series resistance-dependent symmetry that reversibly maps the I-V characteristics of the device.
\end{theorem}
{\em Proof:} To outline the proof, we construct an ordinary differential equation (ODE) from the known solution and then show that this ODE indeed has translational symmetry. We begin by differentiating Eq. \ref{eqn:schottky} w.r.t $x$. Then, after some algebra, we find that:
\begin{eqnarray}
    \nonumber py'&=&be^{c(x-yr)}e^{-b(x-yr)}-bre^{c(x-yr)}e^{-b(x-yr)}y'\\
    &+&c\Big\{e^{c(x-yr)}-e^{c(x-yr)}e^{-b(x-yr)}\Big\}(1-ry'),\label{eqn:schottky2}
\end{eqnarray}
where $y'$=$dy/dx$. Rearranging Eq. \ref{eqn:schottky2} gives the Schottky diode equations as a first-order ODE:
\begin{equation}
    y'=\frac{\beta(x,y)}{p+r\beta(x,y)}=\omega(x,y),
    \label{eqn:schottky3}
\end{equation}
where $\beta(x,y)$
\begin{equation}
    \beta=\beta(x,y)=\Big[(b-c)e^{-b(x-yr)}+c\Big]e^{c(x-yr)}.
    \label{eqn:beta1}
\end{equation}
Eq. \ref{eqn:schottky3} denotes the instantaneous diode conductance at a given ($x$,$y$) point. Since a localized gradient can be estimated with good accuracy from an experimental data plot near any point, this result provides an avenue to exploit Eq. \ref{eqn:schottky3} experimentally. The conductance also depends on the series resistance $r$ of the device, on absolute diode temperature (through $b$ and $c$), and on $p$=$p(A,A^*,b,\Phi_0)$.    
Devices with zero internal resistance are the special case i.e. $r$=0, so that $y'$=$\beta/p$, where $\beta$ is itself a simpler expression. Needless to say, such devices are neither practical nor typical. We therefore focus on the general case where $r$$\ne$0.

Our interest is to find the non-trivial symmetries i.e. those that do not map the solution ($x,y$) points onto themselves. In general, a transformation is a Lie symmetry if it preserves the structure of the Lie group of the ODE. Stated differently, there is a symmetry if the closure property holds i.e. an action of the group, such as differentiation, produces elements that are still in the group. Specifically for Eq. \ref{eqn:schottky3}, if a symmetry exists and, if the action is such that ($x,y$) $\mapsto$ ($\hat{x},\hat{y}$), then the transformed ODE should be writable as 
\begin{equation}
    \hat{y}'=\frac{\hat{\beta}}{p+r\hat{\beta}},
    \label{eqn:schottky4}
\end{equation}
where $\hat{\beta}$=$\hat{\beta}(\hat{x},\hat{y})$=$\beta(\hat{x},\hat{y})$. In general, a given ODE may have any number of symmetries that satisfy the closure property. A parametric plot of the known solutions can give clues about the allowed transformations that make it a true Lie symmetry. This is useful not only to speed up the search for symmetries, but to highlight specific domains of validity in ODEs that may have singular solutions. Returning to the specific case of Fig. \ref{fig:symmetry}, the apparent translational symmetry suggests that the transformations have the form:
\begin{equation}
    \hat{x}=x+\mu\epsilon,\quad\textrm{and}\quad\hat{y}=y+\delta\epsilon,
    \label{eqn:transform1}
\end{equation}
where $\epsilon$ is the infinitesimal variational parameter, and $\mu$ and $\delta$ are determinable constants that somehow differentiate any two curves that may arise as a result of varying some inherent parameter. Substituting Eq. \ref{eqn:transform1} into Eq. \ref{eqn:beta1} leads to 
\begin{equation}
    \hat{\beta}=\Big[(b-c)e^{-b(\hat{x}-\hat{y}r)}e^{b(\mu\epsilon-\delta\epsilon r)}+c\Big]e^{c(\hat{x}-\hat{y}r)}e^{c(\mu\epsilon-\delta\epsilon r)}.
    \label{eqn:beta2}
\end{equation}
Comparing Eqs. \ref{eqn:beta1} and \ref{eqn:beta2} shows that the structure of $\beta$ is preserved if and only if $\mu$=$\delta r$. Hence 
\begin{equation}
    \hat{x}=x+\delta r\epsilon,\quad\textrm{and}\quad\hat{y}=y+\delta\epsilon.
    \label{eqn:transform2}
\end{equation}
Therefore, $\mu$ and $\delta$ must depend linearly on diode series resistance. Also, the condition $\hat{y}'$=$y'$ still holds by differentiation w.r.t $x$ in both spaces. Finally, one can check that Eq. \ref{eqn:schottky4} arises by substituting $x$=($\hat{x}-\delta r\epsilon$) and $y$=($\hat{y}-\delta\epsilon$) into Eq. \ref{eqn:schottky3}, thereby completing the proof. 

\begin{corollary}\label{corol:thm1}
For a given Schottky diode with two parametrically different characteristics it is possible to find a common resistance point ($x$,$y$) that validates Theorem \ref{thm:first}. 
\end{corollary}
The proof of the corollary is implicit. We notice first that no two such characteristics generally overlap. Second, as $r$ is inherent to the device, it will be common in both characteristics. Therefore, a point can be found on each characteristic with the identical $y'$, that satisfies Theorem \ref{thm:first}. In short, for a given device it is possible to map one characteristic to another using the above transformations on the same device.

MS device parameters are known to depend on factors such as applied bias and junction temperature \cite{durmus2011,shili2013}. The complexity of the MS diode equation is the foremost reason for the existence  of the many extraction methods and their varied simplifying assumptions. A common approach selectively restricts the bias range. Further difficulties can be due to some functional interdependence of unknown form in the parameters themselves. Thus the reported barrier height for a given device can differ substantially depending on the method used. The Gaussian distribution is typically applied to account for the spread of values over applied bias or temperature \cite{tasc2018}. Another issue is that results from different sources or methods can exhibit conflicting trends in the behavior of a variable. For instance, whereas one method might suggest that series resistance increases with bias, another might suggest that it decreases \cite{mik2001}. Therefore, a unifying approach is needed to consistently calculate the parameters and with low-variance thus implying a desirable repeatability. 

The above discussion shows that the only requirement of the symmetry method is the existence of a well-defined group whose membership are the MS device characteristics and the allowed operations. With reference to Fig. \ref{fig:symmetry}, our discussion requires that the device resistance $r$ be necessarily constant for Curve 1 and Curve 2 in the {\em vicinity} of $\epsilon$ for the associated ODEs, Eq. \ref{eqn:schottky3} and Eq. \ref{eqn:schottky3d}, to be valid. Using Theorem \ref{thm:first} and its corollary allows us to propose the idea of conservation of resistance in the presence of symmetry in the characteristics. We will define them here as the empirically determined I-V characteristic curves that have the same resistance for a single MS device, but are differentiated apart by the variation of other parameters such as $T$ and $n$. Thus we could consider the general case when Curve 1 $\neq$ Curve 2, and the specific case when Curve 1 $=$ Curve 2. These possibilities are further discussed later in this article. We start at the necessary coordinate transformation, which means moving from the original I-V characteristic to a resistance-compensated I-V characteristic. The small variational parameter above ($\epsilon$ $\rightarrow$ 0) allows near-identity transformations on the ($x,y$) plane: 
\begin{eqnarray}
    \nonumber\hat{x}&=&x+\epsilon\xi(x,y)+\mathcal{O}(\epsilon^2)\\
    \hat{y}&=&y+\epsilon\eta(x,y)+\mathcal{O}(\epsilon^2),\label{eqn:linearized}
\end{eqnarray}
which, in effect, implements a linearization in terms of the tangent vectors:
\begin{equation}
    \xi(x,y)=\frac{d\hat{x}}{d\epsilon}\Big|_{\epsilon=0},\quad\textrm{and}\quad\eta(x,y)=\frac{d\hat{y}}{d\epsilon}\Big|_{\epsilon=0}.
    \label{eqn:transform3}
\end{equation}
It is helpful to recognize $\epsilon$ as the analogy of time ($t$) in Newtonian mechanics. In that case, $x$ and $y$ respectively denote the displacement and the velocity. If evaluated strictly at $t$=$0$, then the vector field ($\xi$, $\eta$) would denote the instantaneous velocity and acceleration, respectively. However, if $t$ gets very small but is non-zero i.e. $t$$\rightarrow$$0$, then ($\xi$, $\eta$) represents the averaged displacement, velocity and acceleration in the vicinity of $t$. We return to this point in a later in the discussion.

In the presence of symmetry, the variation of $\epsilon$ describes all possible orbits s.t. ($x,y$) $\mapsto$ ($\hat{x},\hat{y}$). Eq. \ref{eqn:transform2} implies that $\xi(x,y)$=$\delta r$, and $\eta(x,y)$=$\delta$. The mapping can be expressed instead in terms of some characteristic function $Q$=$Q(x,y, y')$ s.t.
\begin{eqnarray}
    \nonumber\hat{x}=x,\quad&&\hat{y}=y+\epsilon Q(x,y, y')+\mathcal{O}(\epsilon^2),\quad\textrm{and}\\
    Q(x,y,y')&=&\eta(x,y)-\xi(x,y)y'.
    \label{eqn:transform4}
\end{eqnarray}
The trivial mapping ($\hat{x},\hat{y}$)=($x,y$) results when $Q$=0. The trivial mapping is not a particularly useful one since it gives the same solution point and is, therefore, to be excluded generally. We intend to transform the ($x,y$) coordinate system with its ODE to a canonical system in which the  ODE is possibly easier to solve through some inherent symmetry that may exist. For instance, we can define a constrained new system ($f(x,y)$,$g(x,y)$) where the closure property is still met and the group structure remains intact. That is, if $x$=$x(f,g)$ and $y$=$y(f,g)$, then $\hat{x}$=$\hat{x}(\hat{f},\hat{g})$, and $\hat{y}$=$\hat{y}(\hat{f},\hat{g})$. One way to do this is to define a condition whereby all the possible orbits have the same tangent vector at each point, i.e. ($\xi$,$\eta$)=(0,1). This constraint then leads to a canonical system of equations:
\begin{eqnarray}
    \nonumber \xi(x,y)\frac{\partial}{\partial x}f(x,y)&+&\eta(x,y)\frac{\partial}{\partial y}f(x,y)=0\\
    \xi(x,y)\frac{\partial}{\partial x}g(x,y)&+&\eta(x,y)\frac{\partial}{\partial y}g(x,y)=1.
    \label{eqn:canonical1}
\end{eqnarray}
This condition may be met by several functions $f$ and $g$ since canonical coordinates are generally not unique. It is convenient to choose the simplest $f$ and $g$ that one can think of. Consequently,
\begin{eqnarray}
    \nonumber r\frac{\partial}{\partial x}f(x,y)&+&\frac{\partial}{\partial y}f(x,y)=0\\
     r\frac{\partial}{\partial x}g(x,y)&+&\frac{\partial}{\partial y}g(x,y)=\frac{1}{\delta},
    \label{eqn:canonical2}
\end{eqnarray}

Further constraints are that the transformation is invertible. That is, Curve 1 $\rightleftarrows$ Curve 2. Additionally, to have a determinate system of equations in Eq. \ref{eqn:canonical1}, the tangent vectors may not be zero simultaneously. Applying the chain rule on Eq. \ref{eqn:transform1} gives the equivalent derivative operator, $D_x$, that still meets the closure property in the new system i.e
\begin{equation}
    \frac{d\hat{y}}{d\hat{x}}=\frac{D_x \hat{y}}{D_x\hat{x}}=\frac{D_x f}{D_x g}.
    \label{eqn:totderivop}
\end{equation}
where $D_x$=($\partial_x+y'\partial_y$) in first-order, and $\mathcal{O}(\epsilon^2)$. The new system is potentially easier to solve since the ODE becomes invariant along one axis thus reducing the quadrature i.e. the necessary integration. The solution of the ODE is obtained by solving the characteristic equation, $Q(x,y)$=($\eta$-$\xi y'$)=0 i.e.
\begin{equation}
    \frac{dy}{dx}=\frac{\eta(x,y)}{\xi(x,y)}=\frac{\delta}{\delta r}=\frac{1}{r},
    \label{eqn:reducedODE}
\end{equation}
for $r$$\neq$$0$.  In the context of Noether's first theorem mentioned above, Eq.\ref{eqn:reducedODE} implies that the series resistance is a conserved quantity in this symmetry. That is, the characteristics exhibit the same resistance $r$ in the vicinity of $\epsilon$. This resistance can vary with some external parameter e.g. illumination intensity, temperature, doping and so on, but will do so identically for all the characteristics around $\epsilon$. Eq. \ref{eqn:reducedODE} has the general solution $y$=($x/r$+$m$), where $m$ is an arbitrary constant. The ODE in Eq. \ref{eqn:reducedODE} is still physically meaningful since $dy/dx$ denotes the reciprocal of resistance (conductance), $y$ is the current, and $x$ is the applied voltage bias. Also, there are no singular solutions since $r$$>$$0$ for real devices. We can arbitrarily choose a canonical coordinate ($f$, $g$) in terms of $m$, e.g. 
\begin{equation}
    m=f=y-\frac{x}{r},\label{eqn:generalsoln}
\end{equation}
This selection satisfies the first expression in Eq. \ref{eqn:canonical2} and $f$ is invariant to the transformations. Then
\begin{equation}
    g(f,x)=\int\frac{dx}{\xi(y,r)}\Big|_r=\int\frac{dx}{\delta r}=\frac{x}{\delta r}.
\end{equation}
Therefore, a possible canonical coordinate is
\begin{equation}
    (f,g)=\Big(y-\frac{x}{r},~\frac{x}{\delta r}\Big), \quad\textrm{or}\quad (x,y)=\Big(g\delta r,~f+g\delta\Big).
    \label{eqn:newsystem}
\end{equation}
Eq. \ref{eqn:totderivop} allows us to write the transformed derivative:
\begin{equation}
    \frac{dg}{df}=\frac{g_x+\hat{y}'g_y}{f_x+\hat{y}'f_y}=\frac{(1/\delta r)}{(-1/r)+\hat{y}'}.
    \label{eqn:transformedSpace}
\end{equation}
Substituting $x$ and $y$ from Eq. \ref{eqn:newsystem} into Eqs. \ref{eqn:beta1} and \ref{eqn:schottky4} and simplifying the resulting expressions gives $\hat{\beta}$ and $\hat{y}'$ in terms of only one canonical coordinate i.e.
\begin{eqnarray}
    \hat{\beta}&=&\Big[(b-c)e^{-brf}+c\Big]e^{crf},\\
    \hat{y}'&=&\frac{\Big[(b-c)e^{-brf}+c\Big]e^{crf}}{p+\Big[(b-c)e^{-brf}+c\Big]e^{crf}}.\label{eqn:beta3}
\end{eqnarray}
Substituting Eq. \ref{eqn:beta3} into Eq. \ref{eqn:transformedSpace} and simplifying then gives the transformed ODE in only one canonical variable:
\begin{equation}
    \frac{dg}{df}=-\frac{1}{\delta}+\Big(-\frac{r}{p\delta}\Big)\Big\{(b-c)e^{brf}+c\Big\}e^{-crf}.
    \label{eqn:transformedODE}
\end{equation}
This result can be integrated easily w.r.t $f$ to give the MS equation in the chosen canonical space:
\begin{equation}
    g(f)=-\frac{f}{\delta}+\frac{1}{p\delta}\Big[1-e^{brf}\Big]e^{-crf}.
    \label{eqn:resultterm}
\end{equation}
The final check is whether the original Eq. \ref{eqn:schottky} can be obtained from Eq. \ref{eqn:resultterm}. We see that the direct substitution of $f$, after multiplying through by $\delta$ and setting $g\delta$=($y-f$) from Eq. \ref{eqn:newsystem} then gives the original equation:
\begin{equation}
y=\frac{1}{p}\Big[1-e^{brf}\Big]e^{-crf},\label{eqn:origeqn}
\end{equation}
remembering that $f$=($y-x/r$).

\subsubsection{Uniqueness of the symmetry}

Next, we address the question of whether the above ODEs have many symmetries, or only one unique symmetry. We start at the ODE in Eq. \ref{eqn:schottky3}:
\begin{eqnarray}
    y'=\omega(x,y).
\end{eqnarray}
One can try different ansatze ({\em pl.} of ansatz) for the tangent vector field ($\xi$,$\eta$). An ansatz is a guessed, smooth function with arbitrary constraints that is sometimes employed to provide insights into a problem that resists known analytical approaches. Imposing the linearization in Eq. \ref{eqn:linearized} on Eq. \ref{eqn:schottky4} implements the so-called linearized symmetry condition (LSC). This condition can speed up the finding of the symmetries and ultimately ease the solving of the ODE.
\begin{theorem}
The I-V characteristics have one, and only one, symmetry. This unique symmetry is denoted by Eq. \ref{eqn:transform2}.
\end{theorem}
{\em Proof:} We prove by logical fallacy that, for any non-trivial ansatz ($\xi(x,y)$,$\eta(x,y)$), we arrive at the same symmetry. The LSC from Eq. \ref{eqn:schottky3} can readily be shown to be
\begin{equation}
    \eta_x+(\eta_y-\xi_x)\omega-\xi_y\omega^2=\xi\omega_x+\eta\omega_y,
    \label{eqn:lsc}
\end{equation}
in terms of partial derivatives of the variables shown as subscripts. Consider the following, broadly defined ansatze \cite{hydon2000symmetry} for the tangent vector field:
\begin{equation}
    \xi=\alpha(x),\quad\eta=\Gamma(x)y+\gamma(x),
    \label{eqn:ansatz1}
\end{equation}
in terms of some smooth, arbitrary functions $\alpha$, $\Gamma$ and $\gamma$ whose exact expressions are not known {\em a priori}. Direct substitution of Eq. \ref{eqn:ansatz1} into Eq. \ref{eqn:lsc} with $\omega(x,y)$ defined through Eqs. \ref{eqn:schottky3} and \ref{eqn:beta1} leads, after some extensive algebra, to 
\begin{equation}
    \gamma'(x)=\alpha'(x)=0,\quad \Gamma=0,\quad \alpha(x)=r\gamma(x).
\end{equation}
Thus both $\alpha$ and $\Gamma$ are found to be constants. Hence ($\xi$,$\eta$)=($r\gamma$, $\gamma$). This is the same result in Eq. \ref{eqn:transform2}. As a further illustration, one can try a more restrictive ansatz \cite{hydon2000symmetry,ocaya2016}:
\begin{equation}
    \xi=c_1x+c_2y+c_3,\quad \eta=c_4x+c_5y+c_6
    \label{eqn:ansatz2}
\end{equation}
where $\{c_1,\ldots, c_6\}$ are arbitrary constants. This ansatz is frequently used to find common symmetries such as rotations, translations, scalings, etc. Following the same approach, and with similar extensive algebra, we again see that $c_3$=$rc_6$ i.e. ($\xi$,$\eta$)=($rc_6$, $c_6$). Therefore, we arrive at the same non-trivial symmetry regardless of the choice of $\xi$ and $\eta$. The reader is invited to try other ansatze to test the uniqueness of this symmetry. The translational symmetry in the I-V curves is thus evidently unique. We now return to Eq. \ref{eqn:schottky1a} to examine the conditions for possible symmetry when $\phi$ depends on bias. 

\subsection{Case II - Variable barrier height}

Based on the many empirical measurements in the literature, the variation of barrier height with bias $x$ is commonly taken to be linear \cite{durmus2011,mik2001} i.e. \begin{equation}
    \phi(x)=\Phi_0+\alpha x,
\end{equation} 
where $\Phi_0$ is the zero-bias barrier height and $\alpha$ is the constant bias-coefficient of barrier height (in $eV$/$V$). Differentiating Eq. \ref{eqn:schottky1a} w.r.t $x$, noting that $\phi'$=$\alpha$, gives:
\begin{eqnarray}
    y'&=&\frac{\Big[(b+\alpha b-c)e^{-b(x-yr)}-(\alpha b-c)\Big]e^{c(x-yr)}}{se^{b\Phi_0}+r\Big[c+(b-c)e^{-b(x-yr)}\Big]e^{c(x-yr)}}    \label{eqn:schottky3c}\\
    \quad\textrm{or}~&& y'=\frac{\omega(x,y)}{p+r\zeta(x,y)},\label{eqn:schottky3d}
\end{eqnarray}
where $p$=$s\exp(b\Phi_0)$. Eq. \ref{eqn:schottky3c} reduces to Eq. \ref{eqn:schottky3} when $\alpha$=0, since in that case we get $\omega$=$\zeta$=$\beta$ again. Furthermore, it is similarly easy to see that the symmetry condition for Eq. \ref{eqn:schottky3c}:
\begin{equation}
    \hat{y}'=\frac{\hat{\omega}(\hat{x},\hat{y})}{p+r\hat{\zeta}(\hat{x},\hat{y})}
\end{equation}
still holds, and is unique only for the transformations in Eq. \ref{eqn:transform2}. This suggests that the results preclude local variations around $\epsilon$ (or more accurately $\Delta\epsilon$). This can be appreciated when, for instance, one assumes that the ideality factor $n$ depends on applied bias in some unknown way \cite{mik1999}. But, evidently, $\Delta x$ does not significantly affect $n$ around $\Delta\epsilon$. This observation allows $n(x)$ to be determined by considering contiguous regions of increasing $x$.

\section{The series resistance}

\begin{figure}
\centering
\includegraphics[width=0.7\textwidth]{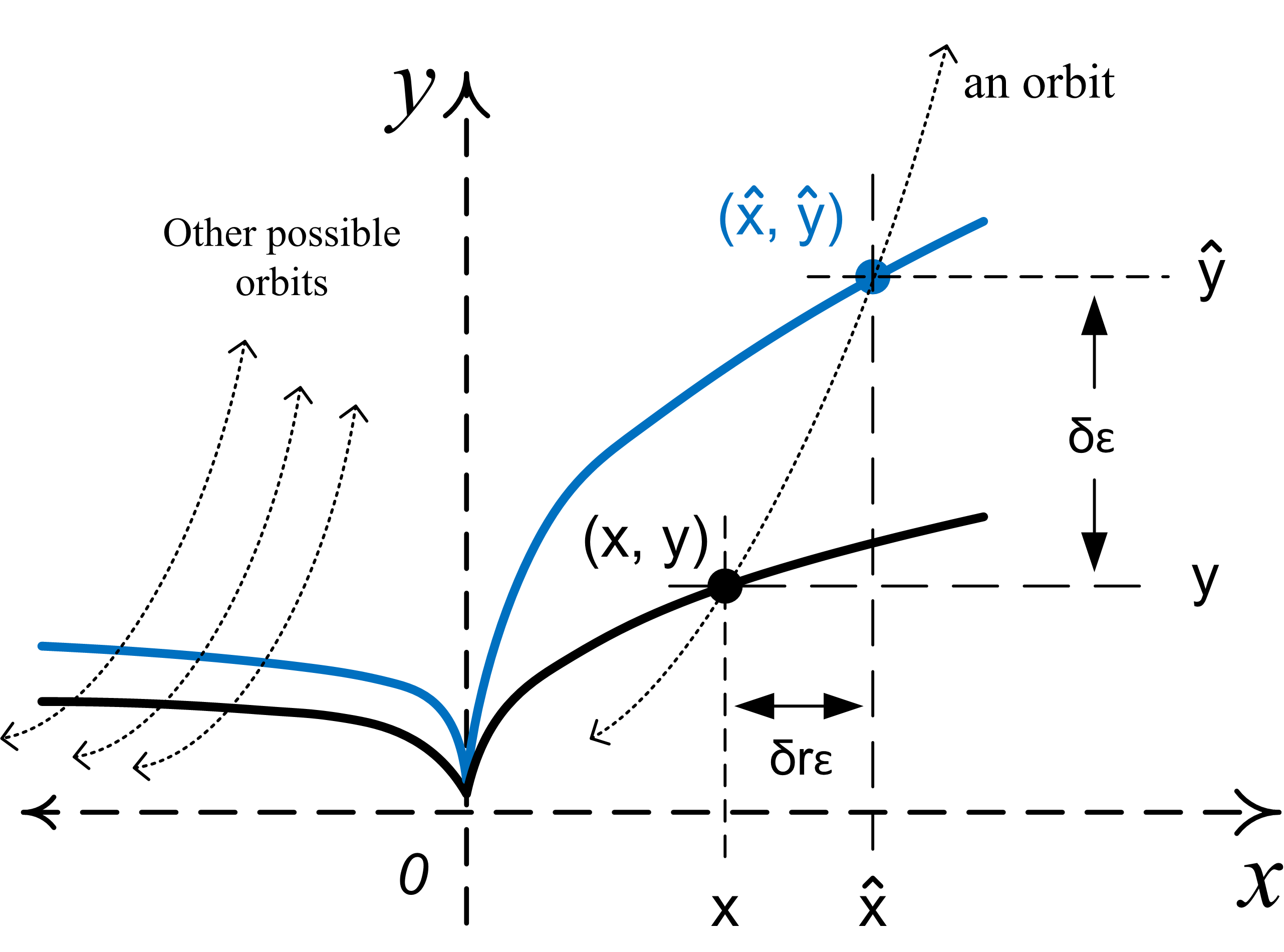}
\caption{\label{fig:thematic}Thematic representation of the symmetry action of Eq. \ref{eqn:transform1} on the points ($x$, $y$) and ($\hat x$, $\hat  y$) that are located on the I-V characteristics of the MS diode. The action is illustrated in the +$x$, forward bias region, but is equally valid in the $-x$, or reverse bias region.}
\end{figure}

Fig. \ref{fig:thematic} illustrates the symmetry action on the I-V characteristics of the MS diode. The action is mathematically applicable in the $\pm x$ bias regions, which is a considerable improvement. 
We now return to the allusion that was made in the context of the Newtonian mechanics. In the event that $\epsilon$$\rightarrow$$0$, then the tangent vector field ($\xi$, $\eta$) represents, respectively, the instantaneous rates of change of applied bias and current w.r.t to an arbitrary $\epsilon$. In the experimentally determined I-V characteristics it is more useful to talk of averaged values in the vicinity of $\epsilon$. In other words, if we set $\delta\epsilon$$\rightarrow$$\Delta\epsilon$ in Eq. \ref{eqn:transform2}, then
\begin{equation}
    \hat{x}=x+r\Delta\epsilon,\quad\textrm{and}\quad\hat{y}=y+\Delta\epsilon.
    \label{eqn:averaged}
\end{equation}
The graphical determination of $r$ becomes trivial on an experimental data set:
\begin{equation}
    r=\frac{\hat{x}-x}{\hat{y}-y}=\frac{r\Delta\epsilon}{\Delta\epsilon}.
    \label{eqn:seriesres}
\end{equation}

The compensation for resistance on an I-V data set requires that we first the locate the points ($x$, $y$) and ($\hat x$, $\hat y$). The separation ($x$-$\hat x$) denotes the voltage step of the calculations. In the experimental validation below, the instrumentation acquires data in voltage bias steps of 0.1V. In the ensuing analysis we arbitrarily choose a small enough ($x-\hat x$) e.g. 0.2V, so that Eq. \ref{eqn:seriesres} effectively spans 3-data points while still meeting the condition of Eq. \ref{eqn:averaged} i.e. $\Delta\varepsilon\rightarrow 0$. The modified bias at $x$ was then calculated from 
\begin{displaymath}
(x-yr)|_x \equiv (V-IR_s)|_V,\quad\textrm{where}~~I=\textrm{max.}\{y,~\hat y\}|_\epsilon.
\end{displaymath}
That is, at each $x$ we generate a datum that meets both Eq. \ref{eqn:generalsoln} and Eq. \ref{eqn:origeqn}, thus giving a new set of points ($f$, $y$) in canonical space. The bigger current in the vicinity of $\varepsilon$ is used to find the maximum compensation to the applied bias at the point. 

\begin{figure}
\centering
\includegraphics[width=0.7\textwidth]{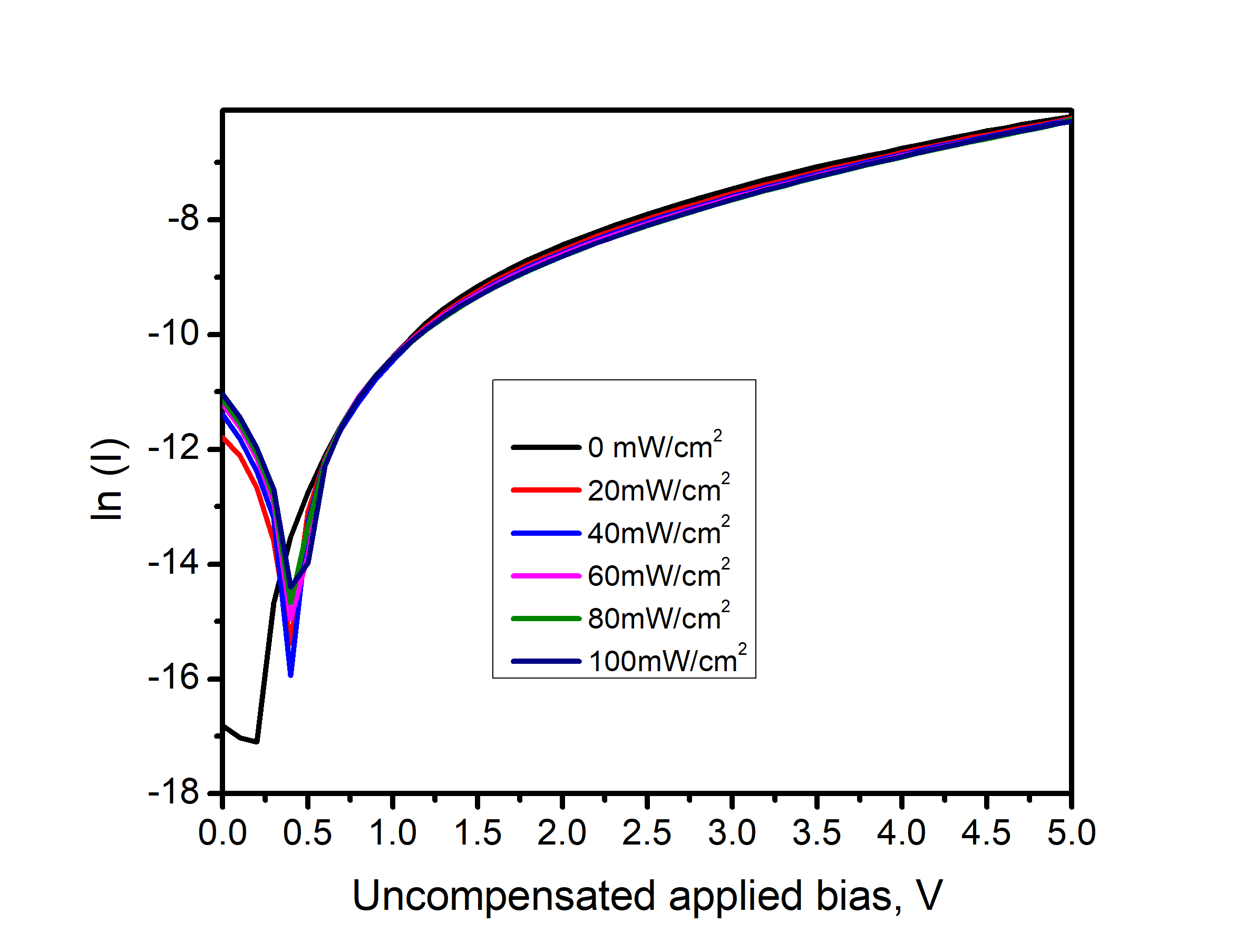}
\caption{\label{fig:uncompIV}Uncompensated semilog I-V characteristics of experimental Al/p-Si/Bi$_2$Se$_3$/Al diode.}
\end{figure}

\begin{figure}
\centering
\includegraphics[width=0.7\textwidth]{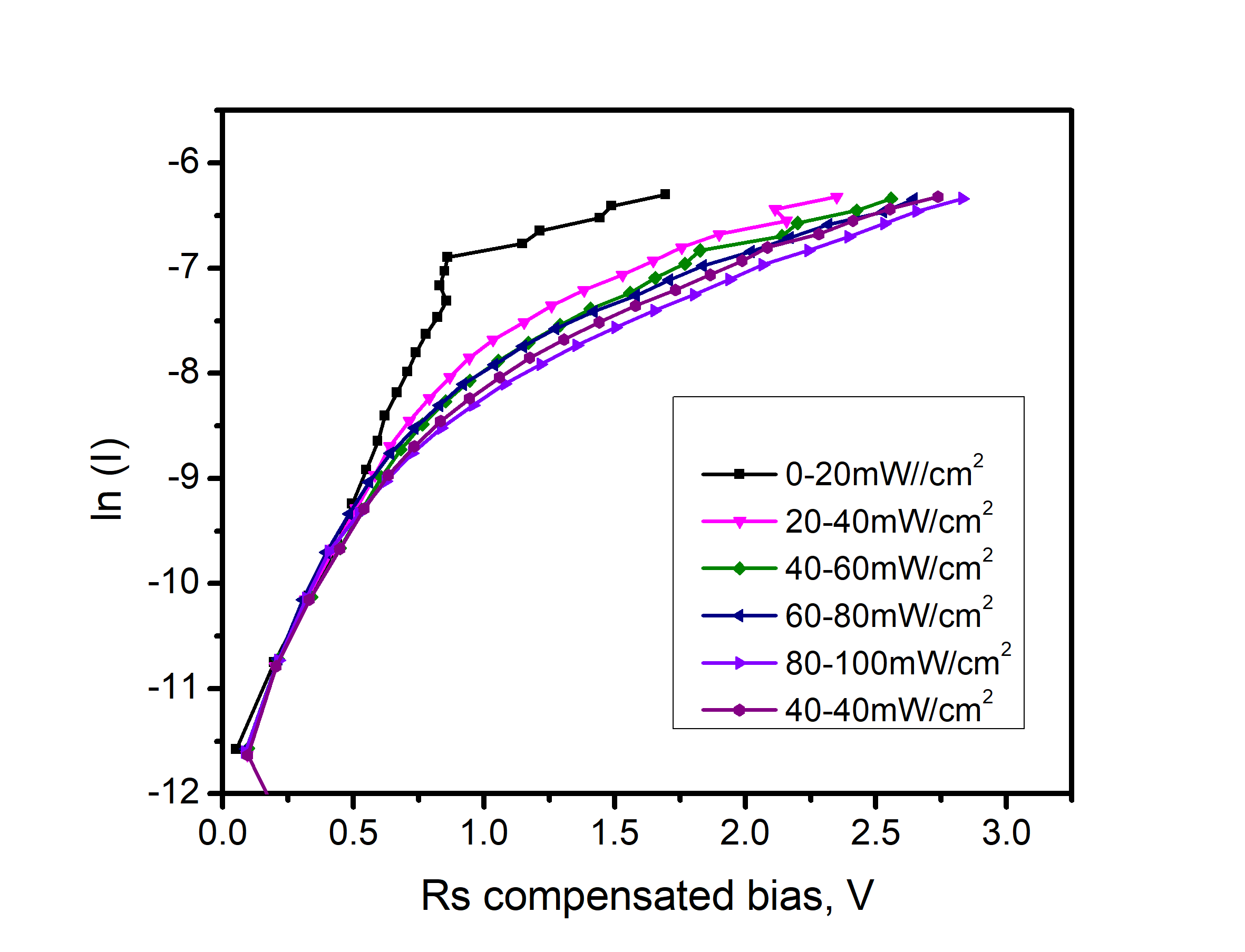}
\caption{\label{fig:compIV}Compensated semilog I-V characteristics of the device under the test cases mentioned in the text. The 40-40 mW/cm$^2$ plot is the test case where Curve 1 = Curve 2.}
\end{figure}

\begin{figure}\centering
\includegraphics[width=\textwidth]{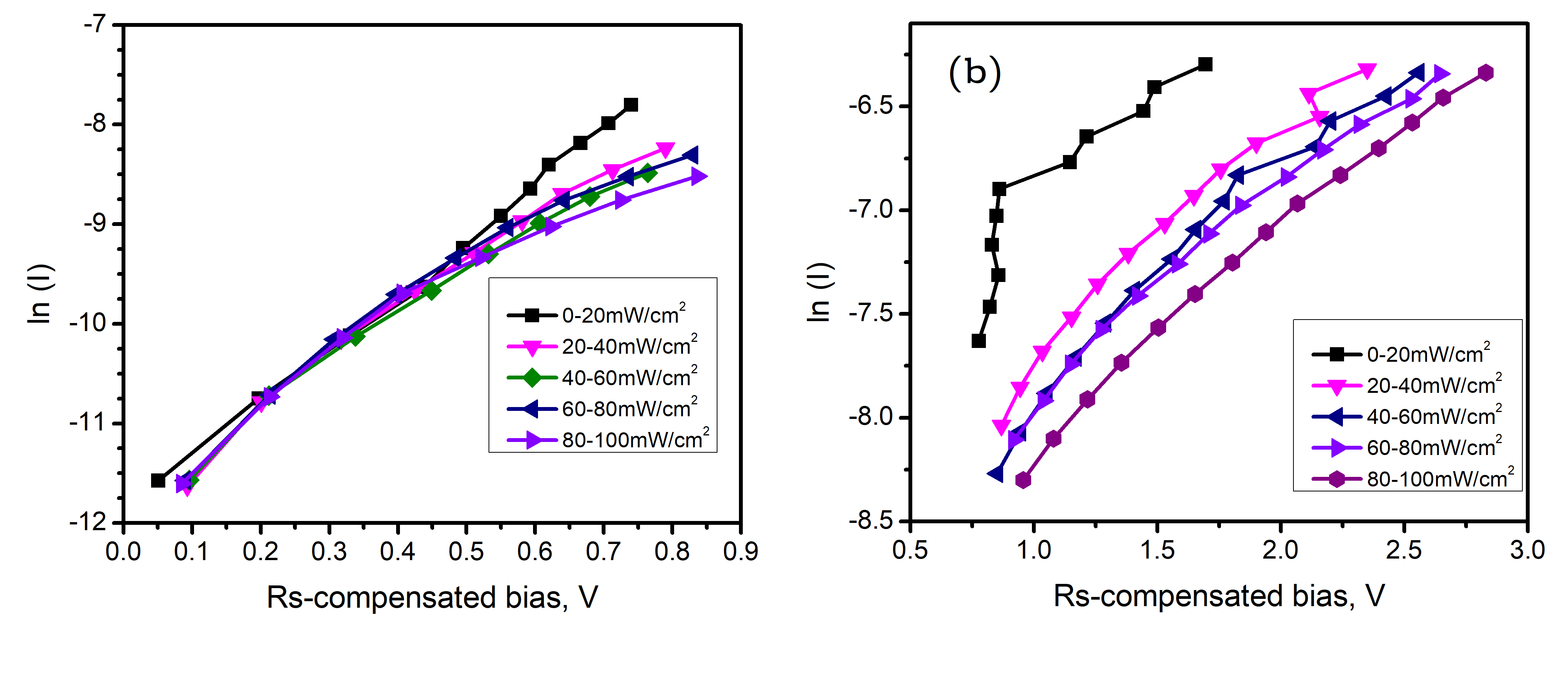}
\caption{\label{fig:lowhighbias}Comparison of the semilog plots in the low and high compensated bias regions. In (a), the plot is generated in part of the reverse bias, as is apparent in Fig. \ref{fig:uncompIV}.}
\end{figure}

\begin{figure}
\centering
\includegraphics[width=0.7\textwidth]{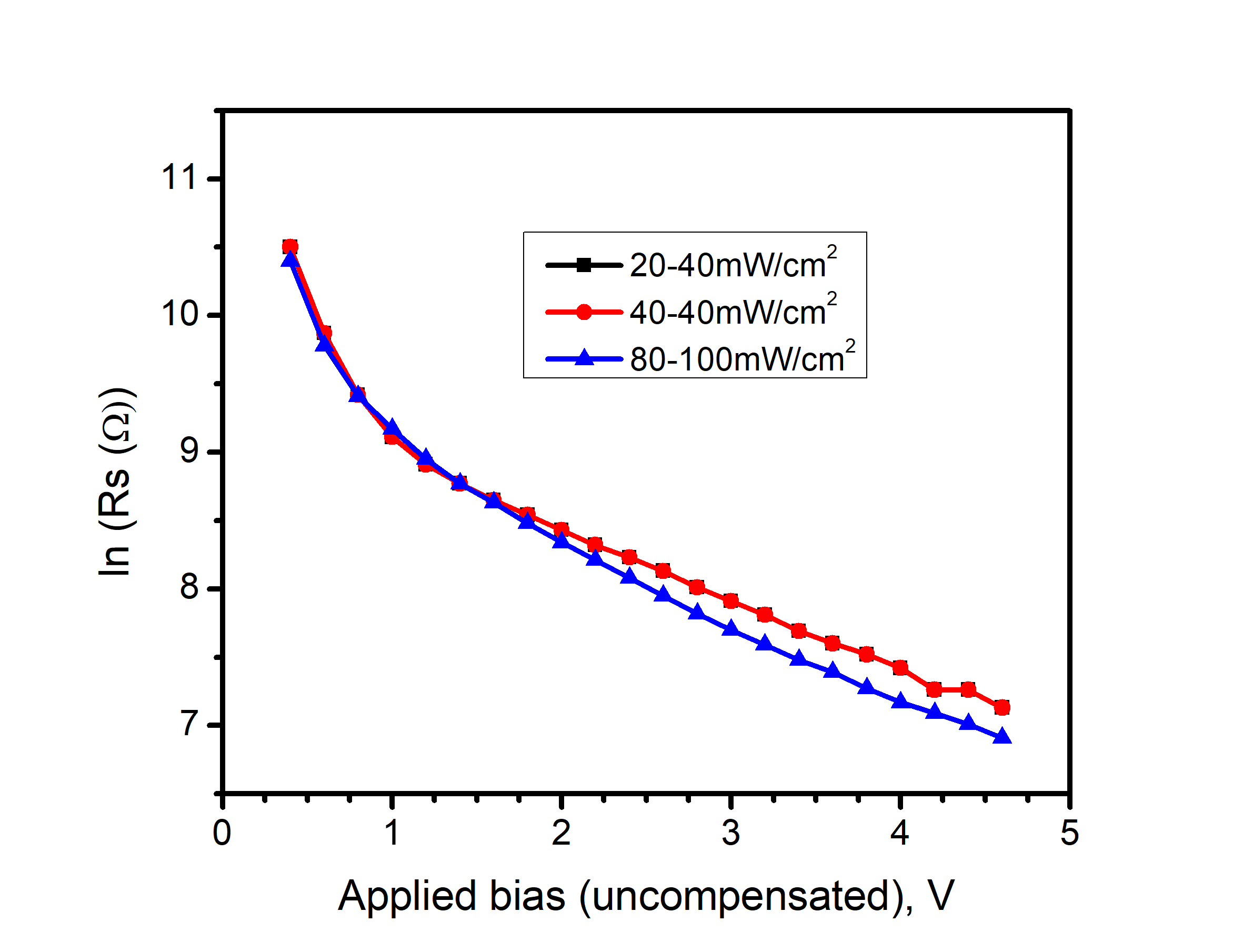}
\caption{\label{fig:Rsymmetry}Plots showing the test cases where Curve 1 $\neq$ Curve 2 (20-40, and 80-100 mW/cm$^2$ ), and Curve 1 = Curve 2 (40-40 mW/cm$^2$). In all cases the resistance at a given bias depends to a small extent on the illumination.}
\end{figure}

Many semilog I-V methods approximate the MS equation by discarding exponential terms. This narrows the bias range of application and gives less reliable parameter estimates. In the absence of a rigorous error analysis the true impact of such approximations is difficult to ascertain. The proposed approach avoids these difficulties. There are two cases for the experimental validation. In the first case the points ($x$, $y$) and ($\hat x$, $\hat y$) are chosen on two different characteristics. In the second case the point ($\hat x$, $\hat y$) is located on the same characteristic as ($x$, $y$) but s.t. $\hat x$$>$$x$. Having determined the canonical data set we then obtain logarithm plot of Eq. \ref{eqn:origeqn}, and compare the results of $R_s$, $n$ and $\Phi_0$ using the method on the two cases. The Cheung-Cheung method \cite{cheung1986extraction} uses the following functions:
\begin{eqnarray}
    \frac{dV}{d\ln I}=R_sI&+&\frac{nkT}{q},~~\textrm{and}~~H(I)=R_sI+n\Phi_b,\\
    \nonumber\textrm{where}&&\\
    ~H(I)=V&-&\frac{nkT}{q}\ln\Big(\frac{I}{AA^*T^2}\Big).
\end{eqnarray}
The symbols have their usual meanings. At a given $T$ the functions $dV$/$d\ln (I)$ and $H(I)$ are plotted linearly against $I$. This gives two estimates of $R_S$. The intercept of the first plot is then used to estimate $n$ and, with the intercept of the second plot, leads to an estimate of $\Phi_b$. The Norde method \cite{norde1979,gromov1994modified} follows a similar approach.
Fig. \ref{fig:uncompIV} and Fig. \ref{fig:compIV} respectively show the raw data and $R_s$ compensated I-V semilog plots of the characteristics under 0 to 100 mW/cm$^2$ illumination. Table \ref{tab:newmethod} shows the results of the method on the Al/p-Si/Bi$_2$Se$_3$/Al diode under the two test cases. For the experimental diode used, $A^*$=32 A/K$^2$cm$^2$, $A$=1mm, and $T$=300K. This allows $\Phi_b$ to be calculated graphically.
Fig. \ref{fig:symmetry} shows that the variation of resistance with applied bias in both bias regions has the form:
\begin{equation}
    R_s=R_0e^{-\alpha V_{a}},
\end{equation}
where $V_a$ is applied bias, and $\alpha$ is a constant. The calculated $R_0$ are 61.4 k$\Omega$ and 13.2 k$\Omega$ respectively. Earlier work based on different Schottky diodes \cite{mekki2016new} support the trends seen in capacitance-voltage impedance measurements i.e. that the series resistance decreases with bias. Our results further suggest that $\alpha$ depends on the illumination and bias region. It increases linearly with illumination in the first region of bias, but decreases linearly in the second region. Similar trends in the resistance versus both applied bias and illumination have been reported by Lapa et. al. \cite{lapa2019effect}. For this diode, the plots show a significant jump in the diode conduction from around 0.476V e.g. Figs. \ref{fig:compIV}--\ref{fig:Rsymmetry}. We suggest that this bias represents the built-in potential of the diode, where strong current conduction begins, marked by a significant change in $R_0$. The barrier height is calculated using Eq. \ref{eqn:origeqn}. Denoting the resistance-compensated bias by $z$, and $w$=($1/p$), Eq. \ref{eqn:origeqn} can be approximated written:
\begin{equation}
    \ln y=cz+\ln w.
\end{equation}
That is, a plot of $\ln y$ versus $z$ in both bias regions is a straight line. The gradient $c$ leads to $n$, since $n$=$b/c$, where $b$=$q/kT$. The intercept leads to
\begin{equation}
    \textrm{intercept}=AA^*T^2e^{-b\Phi_b}.
\end{equation}
Furthermore, we see that the calculations based on Curve 1 = Curve 2 when $P$=40 mW/cm$^2$ are practically identical to Curve 1 $\neq$ Curve 2 in the vicinity of the illumination. This means that the method can be applied with good accuracy to measurements that have only one characteristic curve. Table \ref{tab:cheungcheung} summarises the results of the Cheung-Cheung method on the Al/p-Si/Bi$_2$Se$_3$/Al Schottky photo diode. The method also suggests that the two bias regions influence the series resistance, being generally lower for higher bias. In Table \ref{tab:cheungcheung} both regions show an increase of resistance with illumination. From a physical perspective one expects that the conductance should increase with illumination i.e. the resistance decreases. Fig. \ref{fig:lowhighbias} shows that the semilog plots for the two regions of bias. The calculated $n$ and $\Phi_b$ using the two methods are generally comparable.   

\begin{table}
\caption{\label{tab:newmethod}Calculated $n$ and $\Phi_b$ using the proposed method in the two identified regions. The percent variances in $\Phi_b$ w.r.t average $\Phi_b$ in the first and second bias regions are 0.007 and 0.044, respectively.}

\begin{tabular}{lrrrrrr}
  P  & \multicolumn{2}{c}{Bias (0.1V-1.0V)} &  & \multicolumn{2}{c}{Bias (1.0V-4.8V)} \\
\cline{2-4}\cline{5-7}
(mW/cm$^2$) & $n$ & $\alpha$\footnotemark[3] (/V) & $\Phi_b$ (eV) & $n$ & $\alpha$\footnotemark[4] (/V) & $\Phi_b$ (eV) \\
\hline
 0 - 20 & 7.1 & -2.027 & 0.567 & 47.7 & -0.417 & 0.458 \\
 0 - 100 & 7.3 & ---\footnotemark[1] & 0.566 & --- &--- & ---\\
 20 - 40 & 8.9 & -1.914 & 0.561 & 37.9 & -0.511 & 0.483 \\
 20 - 100 & 5.9 & --- & 0.575 & 43.9 & --- & 0.468 \\
 40 - 60 & 10.3 & -1.832 & 0.557 & 39.9 & -0.530 & 0.486 \\
 60 - 80 & 8.9 & -1.679 & 0.560 & 33.9 & -0.578 & 0.495 \\
 80 - 100 & 9.7 & -1.806 & 0.559 & 37.6 & -0.591 & 0.497 \\
 40 - 40\footnotemark[2] & 10.8 & -1.914 & 0.557 & 39.2 & -0.511 & 0.491 \\
\end{tabular}

\footnotetext[1]{not available}
\footnotetext[2]{test with case Curve 1 = Curve 2.}
\footnotetext[3]{$\alpha= 3.51\times 10^{-3}P-2.03$}
\footnotetext[4]{$\alpha= -2.22\times 10^{-3}P-0.41$}
\end{table}

\begin{figure}
\centering
\includegraphics[width=\textwidth]{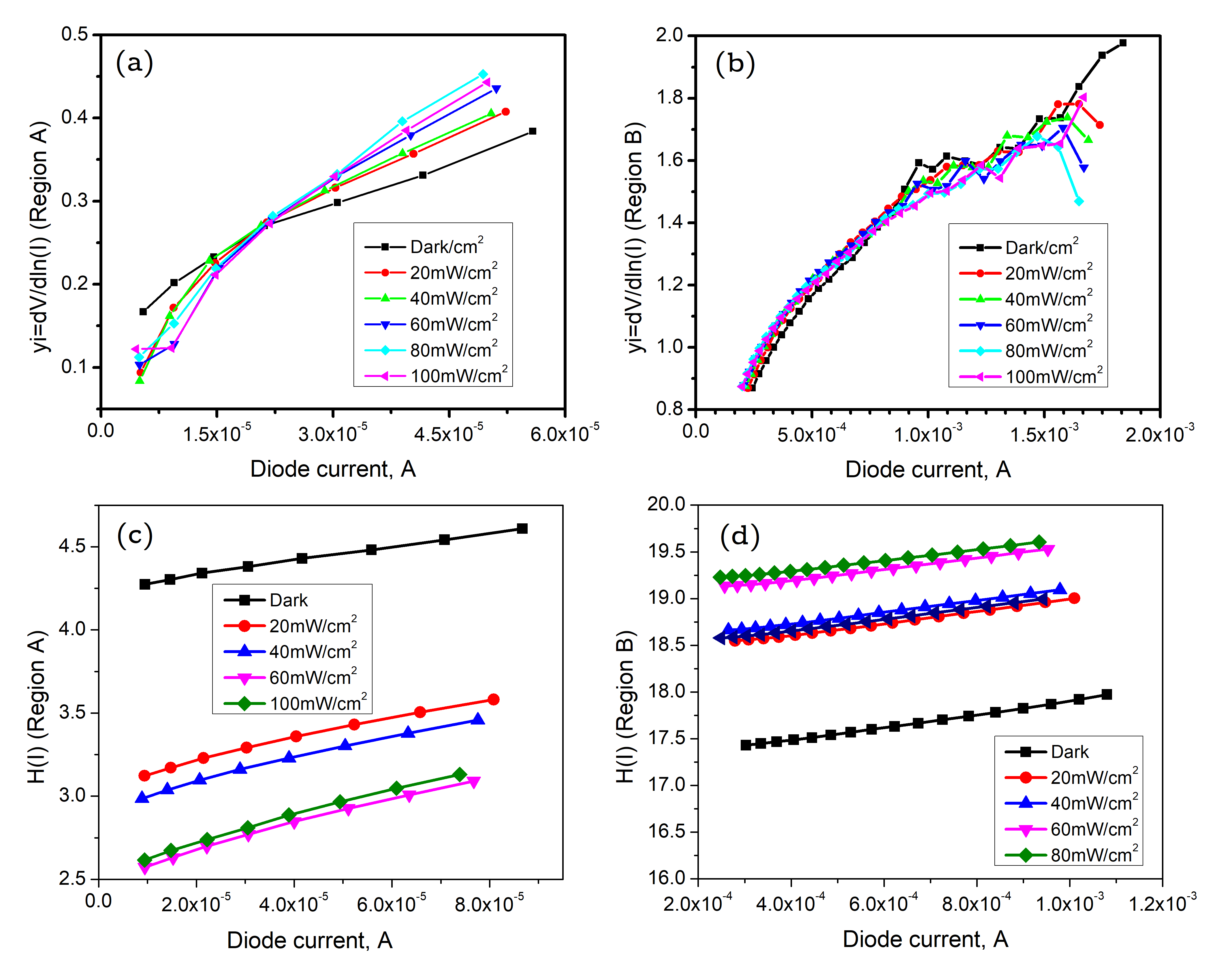}
\caption{\label{fig:dVdlnIregionA}Cheung-Cheung function plots. The bias range is 0.6V-1.40V in (a), (c) and 2.3V-3.9V in (b) and (d). The functions showed low linearity outside these ranges.}
\end{figure}
\begin{landscape}
\begin{table}
\caption{\label{tab:cheungcheung}Calculated $Rs$, $n$ and $\Phi_b$ using the Cheung-Cheung method in the two identified regions. The variances in $\Phi_b$ in percent of average $\Phi_b$ in Regions A and B are 0.088 and 0.004, respectively.}
\begin{tabular}{lllllllllll}
\centering
&\multicolumn{4}{c}{$dV/d\ln (I)$ vs $I$}&\multicolumn{5}{c}{$H(I)$ vs $I$}\\
\cline{2-5}\cline{6-9} 
&\multicolumn{2}{c}{Region A (0.6-1.2V)}&\multicolumn{2}{c}{Region B (2.3-4.8V)} &\multicolumn{2}{c}{Region A (0.6-1.2V)}&\multicolumn{2}{c}{Region B (2.3-4.8V)}&\multicolumn{2}{c}{Averages}\\
\hline
  & n & $R_s(k\Omega)$ & n & $R_s(k\Omega)$ & $R_s(k\Omega)$ & $\Phi_b$ (eV) & $R_s(k\Omega)$ & $\Phi_b$ (eV) & $R_s(k\Omega)$\footnote{Region A, $dV/d\ln (I)$ function} & $R_s(k\Omega)$\footnote{Region B, $H(I)$ function}   \\
\hline
   0 & 6.4 & 4.07 & 32.2 & 0.64 & 3.26 & 0.785 & 0.72 & 0.524 & 3.66 & 0.68 \\
  20 & 4.3 & 6.11 & 34.5 & 0.57 & 6.37 & 0.712 & 0.63 & 0.532 & 6.24 & 0.60 \\
  40 & 4.1 & 6.84 & 34.6 & 0.56 & 6.79 & 0.723 & 0.62 & 0.534 & 6.82 & 0.59 \\
  60 & 3.4 & 7.32 & 35.6 & 0.52 & 7.67 & 0.753 & 0.58 & 0.533 & 7.49 & 0.56 \\
  80 & 3.6 & 7.67 & 35.7 & 0.50 & 7.95 & 0.740 & 0.56 & 0.535 & 7.81 & 0.53 \\
 100 & 3.4 & 7.46 & 34.3 & 0.56 & 7.86 & 0.748 & 0.61 & 0.537 & 7.66 & 0.58 \\
\end{tabular}
\end{table}
\end{landscape}

\section{Conclusions}

In conclusion, we have reported on a previously unseen symmetry in the characteristics of the Schottky diode when it is written as an ODE. We proved that the symmetry is series resistance dependent, translational, and unique. The result is a method that employs symmetry, is simple and directly applies to single or multi-characterisitic I-V plots without undue approximations. Our results on a new Al/p-Si/Bi$_2$Se$_3$/Al diode are compared with the Cheung-Cheung method. The ideality factor and barrier height results are comparable where the bias range of the latter method allowed comparison. Our method gives trends in the parameters that match many literature w.r.t. series resistance, ideality factor and barrier height but with an indication for built-in potential. The approach presented here opens up considerable scope for additional future work. This can involve the search for possible symmetries and actions w.r.t temperature, device doping, and so on.  

\nocite{*}
\bibliographystyle{elsarticle-num}
\bibliography{refs}

\end{document}